\documentclass[lettersize,journal]{IEEEtran}
\usepackage{amsmath,amssymb,amsfonts}
\usepackage{array}
\usepackage[caption=false,font=normalsize,labelfont=sf,textfont=sf]{subfig}
\usepackage{textcomp}
\usepackage{stfloats}
\usepackage{url}
\usepackage{verbatim}
\usepackage{graphicx}
\usepackage{cite}
\usepackage{graphicx}
\usepackage{booktabs}
\usepackage{threeparttable}
\usepackage[table]{xcolor}
\usepackage{tabularx}
\usepackage{hyperref}
\usepackage{xcolor}

\usepackage[linesnumbered,ruled,vlined,nofillcomment]{algorithm2e}
\SetKwInput{KwInit}{Init}
\SetKwInput{KwInput}{Input}
\SetKwProg{Fn}{Function}{:}{}

\begin{document}
\newcolumntype{C}[1]{>{\centering\arraybackslash}p{#1}}

\title{NaviAirway: a Bronchiole-sensitive Deep Learning-based Airway Segmentation Pipeline}

\author{Andong~Wang,
        ~Terence~Chi~Chun~Tam,
        ~Ho~Ming~Poon,
        ~Kun-Chang~Yu,
        ~and~Wei-Ning~Lee
\thanks{This project was in part supported by COVID-19 Action Seed Funding of Faculty of Engineering, The University of Hong Kong. Wei-Ning Lee is the corresponding author.}
\thanks{Andong Wang is with the Department of Electrical and Electronic Engineering, The University of Hong Kong, Hong Kong, China (e-mail:wangad@connect.hku.hk).}
\thanks{Terence Chi Chun Tam is with Respiratory Division, Department of Medicine, The University of Hong Kong, Hong Kong, China, and also with Queen Mary Hospital, Hong Kong, China (e-mail:tcctam@netvigator.com).}
\thanks{Ho Ming Poon is with the Department of Electrical and Electronic Engineering, The University of Hong Kong, Hong Kong, China (e-mail:hmpoon6@connect.hku.hk).}
\thanks{Kun-Chang Yu is with Broncus Medical, Inc., San Jose, CA, 95134 USA (e-mail:jyu@broncus.com).}
\thanks{Wei-Ning Lee is with the Department of Electrical and Electronic Engineering, The University of Hong Kong, Hong Kong, China, and also with the Biomedical Engineering Programme, The University of Hong Kong, Hong Kong, China (e-mail:wnlee@eee.hku.hk).}}

\markboth{}%
{} 

\IEEEpubid{}

\maketitle

\begin{abstract}

Airway segmentation is essential for chest CT image analysis. Different from natural image segmentation, which pursues high pixel-wise accuracy, airway segmentation focuses on topology. The task is challenging not only because of its complex tree-like structure but also the severe pixel imbalance among airway branches of different generations. To tackle the problems, we present a \emph{NaviAirway} method which consists of a bronchiole-sensitive loss function for airway topology preservation and an iterative training strategy for accurate model learning across different airway generations. To supplement the features of airway branches learned by the model, we distill the knowledge from numerous unlabeled chest CT images in a teacher-student manner. Experimental results show that NaviAirway outperforms existing methods, particularly in the identification of higher-generation bronchioles and robustness to new CT scans. Moreover, NaviAirway is general enough to be combined with different backbone models to significantly improve their performance. NaviAirway can generate an airway roadmap for Navigation Bronchoscopy and can also be applied to other scenarios when segmenting fine and long tubular structures in biomedical images. The code is publicly available on \url{https://github.com/AntonotnaWang/NaviAirway}.

\end{abstract}


\section{Introduction}
\label{sec:introduction}
\IEEEPARstart{C}{omputed} Tomography (CT) \cite{deans2007radon} prevails in the assessment of lung diseases, such as lung cancer and chronic obstructive pulmonary disease (COPD). Airway segmentation plays a vital role in the CT image analysis procedure. For example, Navigation Bronchoscopy (NB) is the safest and superior for accessing peripheral pulmonary lesions~\cite{ishiwata2020bronchoscopic}. For better procedural efficiency and patient care, NB requires a pre-planned 3D airway road map that is segmented and reconstructed from CT images. The road map navigates the bronchoscope down into  bronchioles for target nodule sampling~\cite{edell2010navigational, asano2014virtual, kemp2020navigation}. In the case of COPD, airway segmentation from CT images is the key to accurate measurement of the lumen size and wall thickness of each target airway~\cite{berger2005airway}.

\begin{figure}[t]
  \centering
  \includegraphics[width=8cm]{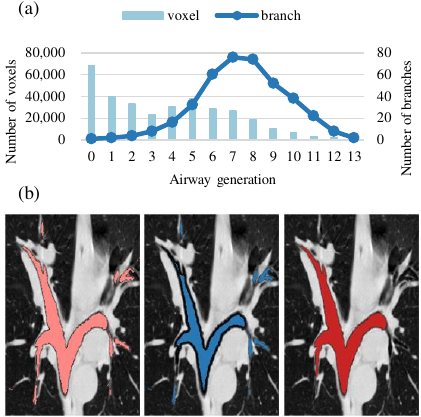}
  \caption{\textbf{(a)} Bar chart: Low-generation airways take up most voxels but have very few branches. \textbf{(b)} The CT image comes from Binary Airway Segmentation (BAS) dataset~\cite{qin2020learning}. From left to right: reference label; the segmentation which has lower dice accuracy but preserves topology; the segmentation which has higher dice accuracy but loses substantial topological information. Our NaviAirway tackles the size imbalance problem while achieving high topological accuracy.}
\label{fig:teaser}
\end{figure}

Airway segmentation is different from natural image segmentation as voxel-wise accuracy is no longer the main concern. Instead, a topologically accurate segment (i.e., preservation of branch connectedness and detection of fine bronchioles) is more important to the success of the aforementioned medical tasks. As shown in Figure~\ref{fig:teaser}, although the red segment has higher voxel-wise accuracy (\textit{e.g.}, dice accuracy), the blue one shows more fine bronchioles and thus provides a better airway road map. In the case of COPD, small airway destruction and narrowing are among the earliest pathological changes, leading to decreased lung function and exacerbation frequency~\cite{agusti2019update}. Therefore, accurate segmentation of fine bronchioles is crucial for early diagnosis and monitoring of COPD.

Besides the morphological complexity, airway segmentation becomes more challenging because of the imbalance sizes among airways of different generations. The airway tree begins from the trachea and ends at the alveoli. The trachea beginning at the larynx is denoted as \emph{Generation 0}, while its subsequently divided left and right main-stem bronchi as \emph{Generation 1}. The airways become progressively finer until the 23rd generation—alveolar sacs~\cite{reece2015overview}. In our paper, \emph{low generation} stands for large airways closer to the trachea, while \emph{high generation} refers to fine bronchioles closer to the alveolar sacs. As shown in Figure~\ref{fig:teaser}, low-generation airways occupy most of the voxels but have fewer branches compared to high-generation airways. Voxel-wise loss functions such as dice loss may lead to the failure of segmenting fine bronchioles. While training on low-generation airways and then on high-generation airways may partly solve the problem ~\cite{zhang2020pathological}, this scheme causes the model to lose knowledge of low-generation airways after long training on high-generation airways.

Over the years, many automated airway segmentation methods have been developed. There are two main categories — traditional methods which rely on manually selected features~\cite{van2013automated, pu2012ct, lo2012extraction, aykac2003segmentation, shi2006upper, cheng2007airway, tschirren2005intrathoracic, tschirren2005segmentation, fabijanska2009two, graham2010robust, fetita2009morphological, kiraly2002three, meng2017automatic, van2008robust} and deep learning-based methods which combine deep learning models with traditional methods or focus on new model architecture design~\cite{cciccek20163d, jin20173d, juarez2018automatic, nadeem2020ct, garcia2021automatic, schlemper2019attention, wang2019tubular, juarez2019joint, selvan2020graph, yun2019improvement, qin2019airwaynet, qin2020airwaynet, zhang2020pathological, qin2021learning, qin2020learning, zheng2021alleviating, wu2021upper, yu2022break, zhang2021fda, gu2022vision, zheng2021refined, zhang2022cfda, wu2022ltsp}. However, effective approaches to guaranteeing topological accuracy and addressing the challenge of imbalanced sizes remain lacking.

Therefore, we present a method, coined as \textbf{NaviAirway}, which is a training framework that can be built on any backbone model. For airway topology preservation, we design a bronchiole-sensitive loss function by pushing the model to recognize fine bronchioles. For accurate model learning across different airway generations, we propose a human-vision-inspired iterative training strategy by guiding the model to specifically extract features of both low- and high-generation airways and preserve those learned features. To further enhance the learning of airway branch features, inspired by Noisy Student~\cite{xie2020self}, we present a teacher-student training method to distill the knowledge from numerous unlabeled CT chest images and increase model accuracy and robustness.


To evaluate model performance, we first test NaviAirway on two public datasets. The comparison results show that our method is more accurate and detects longer airway trees than existing methods. We subsequently test NaviAirway on a private dataset, demonstrating its robustness to a previously unexposed dataset.


Our contributions are summarized as follows:
\begin{itemize}
    \item To our best knowledge, NaviAirway presents the first in-depth study of airway segmentation training framework that focuses on topological correctness, instead of voxel-wise accuracy, and tackles the problems of branch size imbalance. Our method is general, effective, and compatible with any backbone model.
    \item We build a new bronchiole-sensitive loss function that guarantees topological accuracy and drives the model to recognize finer bronchioles (i.e., fine and long tubular shapes), and a new human-vision-inspired iterative training strategy that guides the model to learn both the features of fine and coarse airways while preventing knowledge loss of airway features.
\end{itemize}
\section{Related Works}
\label{sec:related_works}

\subsection{Traditional Methods}
Traditional methods mainly include \textbf{1)} region growing and thresholding, \textbf{2)} morphologic and geometric model-based methods, and \textbf{3)} hybrid approaches combining the above two methods~\cite{van2013automated, pu2012ct}. For example, EXACT’09 Challenge~\cite{lo2012extraction} presented 15 airway tree extraction algorithms submitted to the competition. Ten out of the 15 methods used region growing and thresholding techniques which utilize brightness of different tissues. Similar techniques were also developed, including pixel value filtering and thresholding~\cite{aykac2003segmentation}, thresholding and rectangular region mask~\cite{shi2006upper}, GVF snakes~\cite{cheng2007airway}, fuzzy connectivity~\cite{tschirren2005intrathoracic, tschirren2005segmentation}, and two-pass region growing~\cite{fabijanska2009two}. Alternatively, airways were mathematically defined according to their morphologic and geometric features of airway for extraction~\cite{graham2010robust, fetita2009morphological}. Hybrid methods combined the strengths of the two to provide better segmentation~\cite{kiraly2002three, meng2017automatic, van2008robust}.

\subsection{Deep Learning-based Methods}
Compared with traditional methods, deep learning-based models, on average, detect twice longer airways~\cite{jin20173d, juarez2018automatic, juarez2019joint, qin2019airwaynet, qin2021learning, schlemper2019attention, wang2019tubular, balacey2013mise, nardelli2015optimizing, inoue2013robust, yu2022break, zhang2021fda, gu2022vision, zheng2021refined, zhang2022cfda, wu2022ltsp}. Abundant studies combined Convolutional Neural Networks (CNNs) and traditional methods based on the idea that CNN provided preliminary results, and the traditional method was responsible for refinement. One mainstream of studies used 3D U-Net~\cite{cciccek20163d} as the backbone model and built different post-processing approaches, including fuzzy connectedness region growing and skeletonization guided leakage removal~\cite{jin20173d}, image boundary post-processing to minimize the boundary effect in airway reconstruction~\cite{juarez2018automatic}, and freeze-and-grow propagation~\cite{nadeem2020ct}. Some works focused on designing the backbone network. A simple and low-memory 3D U-Net was proposed in \cite{garcia2021automatic}. Graph Neural Networks (GNNs) were adopted to segment airways~\cite{schlemper2019attention, wang2019tubular}. In \cite{juarez2019joint}, a GNN module was incorporated into a 3D U-Net, while \cite{selvan2020graph} developed a graph refinement-based airway extraction method by combining GNN and mean-field networks. Beyond utilizing existing deep learning models which were built for general tasks, in more recent studies, new network architectures (usually based on 3D U-Net) considering special features of airways were designed to achieve a higher segmentation accuracy. They included patch classification by 2.5 CNN~\cite{yun2019improvement}, AirwayNet~\cite{qin2019airwaynet}, Airway-SE~\cite{qin2020airwaynet}, 2D plus 3D CNN~\cite{zhang2020pathological}, attention distillation modules plus feature recalibration modules~\cite{qin2021learning, qin2020learning}, group supervision plus union loss function~\cite{zheng2021alleviating}, attention on weak feature regions~\cite{wu2021upper}, BREAK~\cite{yu2022break}, and CFDA~\cite{zhang2022cfda}. Despite great progress made by deep learning-based methods, general and effective approaches to airway topology preservation while tackling the problem of size imbalance are lacking.
\section{Method}
\label{sec:method}

\begin{figure*}[t]
  \centering
  \includegraphics[width=18cm]{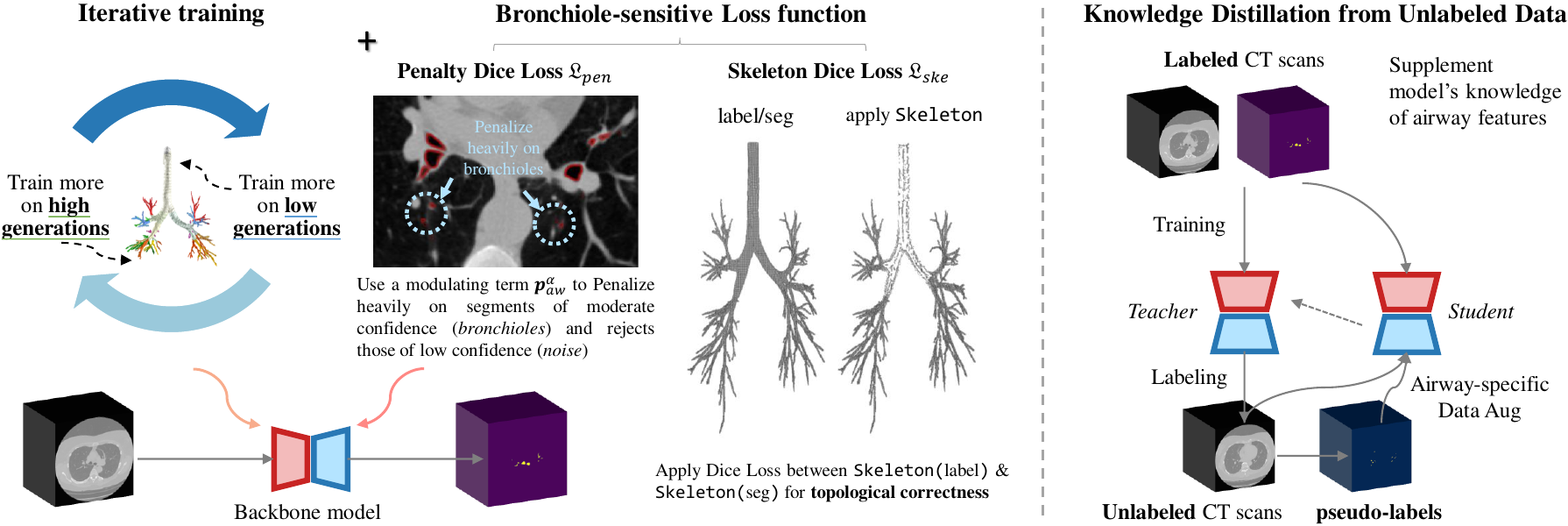}
  \caption{Pipeline of NaviAirway. It consists of a bronchiole-sensitive loss function ($\mathfrak{L}{ske} + \mathfrak{L}{pen}$) that preserves airway topology, and an iterative training strategy that enhances the model's learning across different airway generations. In addition, the model's ability to learn airway features is further improved through knowledge distillation from unlabeled images.}
\label{fig:method}
\end{figure*}

Consider a labeled set $\mathcal{D}: \{(\mbox{\boldmath$x$}_{i}, \mbox{\boldmath$a$}_{i})\}_{i=1}^{N}$, where $\mbox{\boldmath$x$}_{i}$ is a 3D CT image, $\mbox{\boldmath$a$}_{i}$ is the corresponding airway annotation map, and $N$ is the number of labeled images. Note that $\mbox{\boldmath$a$}_{i}$ denotes each voxel on $\mbox{\boldmath$x$}_{i}$ as either background or airway. Also, we have an unlabeled set $\mathcal{U}: \{\mbox{\boldmath$x$}_{i}\}_{i=1}^{M}$, where $M$ is the number of unlabeled 3D CT images and $M \gg N$.

Our pipeline (see Figure~\ref{fig:method}) consists of a backbone model (Section~\ref{sec:method_model}), a new bronchiole-sensitive loss function (Section~\ref{sec:method_loss_func}), and a new human-vision-inspired iterative training strategy (Section~\ref{sec:method_training_strategy}). Then, a teacher-student training technique (Section~\ref{sec:method_distill_knowledge}) further distills the knowledge of airway branch features to increase model robustness. Finally, we introduce our simple post-processing procedure (Section~\ref{sec:method_post_processing}).

\subsection{Backbone Model: Feature Extraction from a Larger Area}
\label{sec:method_model}
Our method offers great flexibility as the backbone model architecture can be customized according to the specific needs of different applications. For our implementation, we designe the model, denoted as $\Phi_{{\theta}}$, based on the well-established 3D U-Net architecture~\cite{cciccek20163d}. We introduce a novel feature extraction module, consisting of one dilated convolution, one self-attention block, and two typical convolutional kernels, to replace the conventional convolution kernels in the down-sampling and up-sampling operations. This innovative module enables the model to extract features from a larger surrounding area, thus preventing interference from other tubular shapes such as the esophagus and vessels, leading to better segmentation results.



\subsection{Loss Function: Topological Correctness}
\label{sec:method_loss_func}

Dice loss~\cite{milletari2016v} is a commonly used loss function for image segmentation that calculates voxel-wise accuracy (see Equation \eqref{equ:org_dice_loss_func}). However, in our case, preserving the topological correctness of the airway is our primary concern. Therefore, we propose our own loss function based on the dice loss

\begin{equation}
\begin{split}
&\FuncSty{Dice}(\mbox{\boldmath$p$}, \mbox{\boldmath$a$}, \mbox{\boldmath$w$})=1-\frac{2\sum_{k=1}^{K}{p_k a_k w_k}}{\sum_{k=1}^{K}{p_k}^2+\sum_{k=1}^{K}{a_k}^2},
\label{equ:org_dice_loss_func}
\end{split}
\end{equation}
\noindent{where} $\mbox{\boldmath$p$}$ denotes the model prediction; $\mbox{\boldmath$a$}$ denotes the reference label; $\mbox{\boldmath$w$}$ is a weight value; $k$ denotes the voxel position; $K$ is the total number of voxels; $p_k \in [0,1]$ is the model confidence of a voxel being background or airway, while $a_k \in \{0,1\}$ is the computer-assisted manual annotation of the $k$-th voxel, where the voxel being the background is assigned as 0 and the voxel being the airway is given a 1.

We use the idea of skeletonization to evaluate topological correctness. Skeletonization is a process which reduces binary images to one-voxel-wide skeletons. If the skeletons of the predicted airway match with the reference labels, the topology is deemed to be successfully preserved. However, current skeletonization methods~\cite{lee1994building} remain a discrete operation, which cannot be applied to a differentiable approximation process. Moreover, the size imbalance among airways of different generations makes skeleton extraction more challenging.

To address this, we present our differentiable semi-skeletonization method, denoted as $\FuncSty{Skeleton}$. We use the morphological operations of erosion and dilation to extract an airway ``skeleton'' that is not strictly one-voxel wide but is enough for topological accuracy calculation. Specifically, we use MinPooling to erode and MaxPooling to dilate (for our implementation, the kernel size $ker = 3$). One iteration of erosion and dilation identifies some fine structures. After multiple iterations (for our implementation, $J = 3$), an airway ``skeleton'' is extracted (Figure~\ref{fig:method}). Our \textbf{Skeleton Dice Loss} $\mathfrak{L}_{ske}$ is formulated as shown in Equation~\eqref{equ:skeleton_dice_loss_func}.

\begin{equation}
\begin{split}
&\mathfrak{L}_{ske} = \FuncSty{Dice}(\FuncSty{Skeleton}(\mbox{\boldmath$p$}_{aw}), \FuncSty{Skeleton}(\mbox{\boldmath$a$}_{aw}), \mbox{\boldmath$1$}),
\label{equ:skeleton_dice_loss_func}
\end{split}
\end{equation}
\noindent{where} subscripts $aw$ denotes the prediction / annotation of airway; $\FuncSty{Skeleton}$ is from Algorithm~\ref{algo:algorithm_skeleton}, and see $\FuncSty{Dice}$ in Equation~\eqref{equ:org_dice_loss_func}.

\begin{algorithm}[t]
\small
\SetKwFunction{FSkeleton}{Skeleton}
\SetKwFunction{FErode}{Erode}
\SetKwFunction{FDilate}{Dilate}
\SetKwFunction{FMaxPool}{MaxPool}
\SetKwFunction{FMinPool}{MinPool}
\SetKwFunction{FReLU}{ReLU}

\KwInput{An image $\mbox{\boldmath$m$}$ in which lower values denote background while higher values denote foreground, kernel size $ker$, number of iterations $J$.}
\KwInit{Skeleton $\mbox{\boldmath$s$}$ $\leftarrow$ a zero image with the same shape as $\mbox{\boldmath$m$}$, padding size $pad \leftarrow \frac{1}{2}(ker - 1)$.}

\Fn{\FSkeleton{\mbox{\boldmath$m$}}}{
    \Repeat{Finishing $J$ iterations} {
        $\hat{\mbox{\boldmath$m$}} \leftarrow \FDilate(\FErode(\mbox{\boldmath$m$}))$ \;
        $\mbox{\boldmath$s$} \leftarrow \FReLU(\mbox{\boldmath$m$} - \hat{\mbox{\boldmath$m$}}) + \mbox{\boldmath$s$}$ \;
        $\mbox{\boldmath$m$} \leftarrow \FErode(\hat{\mbox{\boldmath$m$}})$ \;
    }
    \KwRet $\mbox{\boldmath$s$}$ \;
}

\Fn{\FErode{$\mbox{\boldmath$m$}$}}{
    \KwRet \FMinPool($\mbox{\boldmath$m$}$, $ker$, $pad$) \;
}

\Fn{\FDilate{\mbox{\boldmath$m$}}}{
    \KwRet \FMaxPool($\mbox{\boldmath$m$}$, $ker$, $pad$) \;
}

\caption{Differentiable semi-skeletonization}
\label{algo:algorithm_skeleton}
\end{algorithm}

However, relying solely on Skeleton Dice Loss may not be sufficient as overall accuracy is equally important as topological accuracy. To address this issue, we have designed the \textbf{Penalty Dice Loss} $\mathfrak{L}{pen}$ (see Equation~\eqref{equ:penalty_dice_loss_func}), which includes a modulating term $\mbox{\boldmath$p$}_{aw}^{\alpha}$ to encourage the model to perform better on challenging cases. Since the confidence scores of bronchiole voxels or voxels near the surface between airway branches and background tend to be relatively low, this loss function penalizes the model more heavily for misclassifying such voxels. As the backbone model outputs two prediction maps of the airway and background after the softmax operation, $\mathfrak{L}{pen}$ is composed of two components:

\begin{equation}
\begin{split}
&\mathfrak{L}_{pen} = \mathfrak{L}_{pen}^{aw} + \mathfrak{L}_{pen}^{bg}, \\
&\mathfrak{L}_{pen}^{aw} = \FuncSty{Dice}(\mbox{\boldmath$p$}_{aw}^{\alpha}, \mbox{\boldmath$a$}_{aw}, \mbox{\boldmath$w$}_{aw}), \\
&\mathfrak{L}_{pen}^{bg} = \FuncSty{Dice}(\mbox{\boldmath$p$}_{bg}, \mbox{\boldmath$a$}_{bg}, \mbox{\boldmath$1$}),
\label{equ:penalty_dice_loss_func}
\end{split}
\end{equation}
\noindent{where} subscripts / superscript $bg$ denotes the prediction / annotation of background; $\mbox{\boldmath$w$}_{aw}$ is assigned by the thickness of airway branches; $\alpha$ is an integer and $\mbox{\boldmath$p$}_{aw}^{\alpha}$ is the modulating term representing every element $p_k$ to the power of $\alpha$ (in this paper, $\alpha = 2$).

To address the hard case problem in image segmentation, traditional methods assign larger weights to those difficult cases to improve performance, as shown in Equation~\eqref{equ:org_dice_loss_func}. However, these hand-designed weights are not adaptable during training, which limits their effectiveness. In contrast, we propose a dynamic modulating term, $\mbox{\boldmath$p$}_{aw}^{\alpha}$, which acts as a pseudo-confidence score for difficult cases.

During training, suppose $\alpha$ is 2, for a voxel with $p_k$ = 0.5 with its target score being 1, the resulting $p_{k}^{2}$ value is only 0.25. Therefore, the model is encouraged to predict a $p_k$ value larger than 0.7, resulting in a pseudo-confidence score greater than 0.5 for being an airway. In contrast, during inference, we only use $p_k$ to determine airway segments. This approach is akin to athletes training with additional weights to build strength and then removing them during competition. More analysis and discussion of the Penalty Dice Loss will be presented in Section~\ref{sec:discussion_penalty_dice_loss}.

The final loss function is the sum of $\mathfrak{L}_{ske}$ and $\mathfrak{L}_{pen}$.

\subsection{Iterative Training Strategy: Solution to Size Imbalance}
\label{sec:method_training_strategy}
We address the issue of size imbalance in our segmentation problem, where lower airway generations contain significantly more voxels than higher generations, as shown in Figure~\ref{fig:teaser}. To train our model on this imbalanced data, we first crop the CT images into cuboids and feed them into the model in batches. However, if all cuboids are chosen with the same frequency, the resulting class distribution is severely unbalanced, with few samples for high generations. To overcome this problem, we propose an iterative training strategy inspired by sampling techniques used in class imbalance problems~\cite{he2009learning}.



In our strategy, we adjust the probability of selecting each cuboid pair $(\mbox{\boldmath$x$}_i^j,\ \mbox{\boldmath$a$}i^j)$ in a batch, based on the ratio $r_i^j$ of the number of outermost voxels of the airway segment to the total number of airway voxels, representing the reciprocal of the radius. Larger $r_i^j$ indicates that the airways in $\mbox{\boldmath$a$}i^j$ are fine bronchioles, while smaller $r_i^j$ indicates larger airways. In each iteration, we train the model on both high and low generations, but with different frequencies. Specifically, when the probability of selecting a cuboid pair is proportional to $r_i^j$ (scaled by a temperature parameter $\tau_h$), the model is trained more frequently on high generations (\textit{h iter} in Equation~\eqref{equ:training_strategy}). Conversely, when the probability is inversely proportional to $r_i^j$ (scaled by a temperature parameter $\tau_l$), the model is trained more frequently on low generations (\textit{l iter} in Equation~\eqref{equ:training_strategy}). If $r_i^j = \infty$, indicating no airway exists in the cuboid, we assign a constant probability value $\beta{0}$ (in this paper, $\beta{0} = 1$). We apply a softmax function to normalize the probabilities and use temperature scaling to control the degree of imbalance.

By adjusting the sampling frequencies based on the ratio of outermost airway voxels to the total airway voxels, our iterative training strategy addresses the size imbalance problem in airway segmentation.

\begin{equation}
p_i^j = \left\{ \begin{array}{rcl} {\displaystyle \frac{\exp(r_i^j/\tau_{h})}{{\displaystyle \sum_{r_i^j \neq \infty}\exp(r_i^j/\tau_{h}) + \sum_{r_i^j = \infty}\exp(\beta_{0}/\tau_{h})}}} & (\mbox{\textit{h\ iter}}) \\ {\displaystyle \frac{\exp(1/(\tau_{l}r_i^j))}{{\displaystyle \sum_{r_i^j \neq \infty}\exp(1/(\tau_{l}r_i^j)) + \sum_{r_i^j = \infty}\exp(\beta_{0}/\tau_{l})}}} & (\mbox{\textit{l\ iter}}) \end{array}\right.,
\label{equ:training_strategy}
\end{equation}


\noindent{where} $p_i^j$ represents the probability of the sampled cuboid pair $(\mbox{\boldmath$x$}_i^j,\ \mbox{\boldmath$a$}_i^j)$ in a training batch; $\tau_{h}$ and $\tau_{l}$ are two manually selected temperature values (in this paper, $\tau_{h} = 1$, $\tau_{l} = 0.1$). We will present more analysis of our proposed training strategy in Section~\ref{sec:discussion_training}.



\subsection{Knowledge Distillation from Unlabeled Data}
\label{sec:method_distill_knowledge}

In many cases, we have a larger number of unlabeled images $\mathcal{U}$ than labeled ones $\mathcal{D}$, meaning $M$ is much greater than $N$. To leverage the vast amount of unlabeled data, we propose a teacher-student training framework (inspired by Noisy Student \cite{xie2020self}) to distill the knowledge of the CT chest image distribution \cite{wei2020theoretical} and enhance model accuracy and robustness (see Algorithm~\ref{algo:algorithm_of_teacher_student_training}). Our approach also incorporates airway-specific data augmentation to further enhance the distillation process.

\begin{algorithm}[ht]
\small
\SetKwFunction{FAirwayPseudoLabel}{AirwayPseudoLabel}

\KwInput{An optimized model $\Phi_{{\theta}^\ast}$ trained on labeled dataset $\mathcal{D}$, an unlabeled dataset $\mathcal{U}$ ($M \gg N$), two probability values, $q^t$ and $q^c$, used for data augmentation, number of iterations $I$.}
\KwInit{Teacher model $\Phi_{\theta}^T \leftarrow \Phi_{{\theta}^\ast}$, student model $\Phi_{\theta}^S \leftarrow \Phi_{{\theta}^\ast}$, divide $\mathcal{U}$ into $V$ batches $\left\{\mathcal{U}_b\right\}_{b=1}^V$ (where $V \sim N$).}

\Fn{\FAirwayPseudoLabel{$\mbox{\boldmath$x$}_i$, $q^t$, $q^c$}}{
    $\mbox{\boldmath$A$}_i \leftarrow \Phi_{\theta}^T\left(\mbox{\boldmath$x$}_i\right)>t$ ($t=0.5$ with probability of $q^t$, otherwise $t=0.7$) \;
    $\mbox{\boldmath$A$}_i \leftarrow \mbox{\boldmath$S$}_{h^\ast}$ with probability of $q^c$ (see Sec. \ref{sec:method_post_processing} for the definition of $\mbox{\boldmath$S$}_{h^\ast}$) \;
    \KwRet $\mbox{\boldmath$A$}_i$ \;
}

\Repeat{Finishing $I$ iterations}{
    \For{$\mathcal{U}_b \in \left\{\mathcal{U}_b\right\}_{b=1}^V$}{
        \For{$\mbox{\boldmath$x$}_i \in \mathcal{U}_b$}{
            $\mbox{\boldmath$A$}_i$ $\leftarrow$ \FAirwayPseudoLabel($\mbox{\boldmath$x$}_i$, $q^t$, $q^c$) \;
            Add $\mbox{\boldmath$A$}_i$ to $\mathcal{U}_b$ \;
        }
        Train $\Phi_{\theta}^S$ on $\{\mathcal{D} + \mathcal{U}_b\}$;
        
        
    }
    $\Phi_{\theta}^T \leftarrow \Phi_{\theta}^S$ \;
}
\caption{Distill knowledge from unlabeled data}
\label{algo:algorithm_of_teacher_student_training}
\end{algorithm}

\subsection{Post Processing}
\label{sec:method_post_processing}

This step aims at identifying and removing any disconnected noise shapes. After model training, we obtain an airway confidence map $\Phi_{{\theta}}\left(\mbox{\boldmath$x$}\right)$. By applying a threshold value $th$, we can create an airway mask $\mbox{\boldmath$A$}$ where $\mbox{\boldmath$A$}=\Phi_{{\theta}}\left(\mbox{\boldmath$x$}\right)>th$. Since airway branches are typically connected, we can assume that the largest connected shape in the output is the airway segmentation. Therefore, we identify the largest connected shape $\mbox{\boldmath$S$}_{h^\ast}=\max{\left|\mbox{\boldmath$S$}_h\right|}$ from the set of separated shapes $\left\{\mbox{\boldmath$S$}_h\right\}$ in $\mbox{\boldmath$A$}$. However, there may be some broken airway branches in the other separated shapes $\left\{\mbox{\boldmath$S$}_h\right\} \backslash \mbox{\boldmath$S$}_{h^\ast}$. We connect those broken branches to $\mbox{\boldmath$S$}_{h^\ast}$ if they are close enough. To connect, we first define a search range $R$. For every end point $e$ of airway branches in $\mbox{\boldmath$S$}_{h^\ast}$, if any shape $\mbox{\boldmath$S$}_{\hat{h}}$ in $\left\{\mbox{\boldmath$S$}_h\right\} \backslash \mbox{\boldmath$S$}_{h^\ast}$ is within ${B}_R\left(e\right)$ (i.e., the neighborhood of $e$), update $\mbox{\boldmath$S$}_{h^\ast}$ by $\mbox{\boldmath$S$}_{h^\ast} := \mbox{\boldmath$S$}_{h^\ast}+\mbox{\boldmath$S$}_{\hat{h}}$, and make $\mbox{\boldmath$S$}_{h^\ast}$ a connected shape by lowering threshold $th$ within ${B}_R\left(e\right)$ to fill the gap. In our experiments, $R$ is set to be small (e.g., $R=2$ in this paper) as the broken branch connection is only for refinement purposes.
\section{Experiments}
\label{sec:experiments}

\subsection{Datasets}
Three public datasets (\emph{EXACT’09}~\cite{lo2012extraction},  \emph{Binary Airway Segmentation (BAS)}~\cite{qin2020learning}, and \emph{Airway Tree Modeling Challenge (ATM22)}~\cite{zheng2021alleviating, zhang2021fda, yu2022break, qin2019airwaynet}) and one private dataset (\emph{QMH}) were used to evaluate NaviAirway.
\textbf{1) EXACT’09.} It contains 20 CT images for training and 20 images for testing. The images have slice thickness ranging from 0.45mm to 1.0mm while the pixel spacing ranges from 0.5mm to 0.78mm. \textbf{2) BAS.} It has 90 images (20 from the training set of EXACT’09~\cite{lo2012extraction} and 70 from LIDC-IDRI~\cite{zheng2021alleviating, armato2011lung}) associated with 90 corresponding manually-labeled annotations. The pixel spacing ranges from 0.5mm to 0.8mm, and the slice thickness ranges from 0.5mm to 1.0mm. We follow the instructions in \cite{yu2022break} to split BAS into a training set, a validation set, and a test set. \textbf{3) ATM22.} The 300 training cases (with labels) and 50 validation cases (without labels) are publicly available. All the CT scans were selected from LIDC-IDRI\cite{armato2011lung} and the Shanghai Chest hospital and were labeled by deep learning models and radiologists' manual correction.
\textbf{4) QMH.} Nine cases with slice thicknesses ranging from 1.0mm to 5.0mm and pixel spacings ranging from 0.5 to 0.9 were labeled using commercial software named LungPoint with experts’ manual correction. They were used for external testing because the data distribution was unseen by the model.

There were also a large number of unlabeled cases in LIDC-IDRI (which has 1018 cases in total) and QMH (which has 101 unlabeled cases). Those unlabeled data were used for knowledge distillation of airway branch features to increase model robustness.

\begin{table}[t]
    \setlength{\abovecaptionskip}{0pt}
    \setlength{\belowcaptionskip}{0pt}
    \caption{Performance comparison on BAS dataset. \\ Mean ± standard deviation (\%) is shown for each metric.}
    \centering
    \begin{threeparttable}
    \footnotesize
    \begin{tabular}{l|cccc}
    \toprule
    & DSC & Sensitivity & BD & TD \\
    \cmidrule(lr){1-5}
    Jin et al.~\cite{jin20173d} & 93.6±2.0 & 88.1±8.5 & 93.1±7.9 & 84.8±9.9 \\
    Juarez et al.~\cite{juarez2018automatic} & 93.6±2.2 & 86.7±9.1 & 91.9±9.2 & 80.7±11.3 \\
    AG U-Net~\cite{schlemper2019attention} & 82.7±22.2 & 72.5±28.9 & 70.1±33.3 & 63.5±30.8 \\
    Wang et al.~\cite{wang2019tubular} & 93.5±2.2 & 88.6±8.8 & 93.4±8.0 & 85.6±9.9 \\
    Juarez et al.~\cite{juarez2019joint} & 87.5±13.2 & 77.5±15.5 & 77.5±20.9 & 66.0±20.4 \\
    AirwayNet~\cite{qin2019airwaynet} & \underline{93.7±1.9} & 87.2±8.9 & 91.6±8.3 & 82.1±10.9 \\
    Xue et al.~\cite{xue2020shape} & 92.1±2.4 & - & 87.7±8.1 & 88.2±6.9 \\
    Qin et al.~\cite{qin2020learning} & 91.5±2.9 & - & 87.6±9.2 & 91.8±5.3 \\
    Qin et al.~\cite{qin2021learning} & 92.5±2.0 & 93.6±5.0 & \textbf{96.2±5.8} & 90.7±6.9 \\
    Zheng et al.~\cite{zheng2021alleviating} & 91.4±3.3 & - & 88.7±7.9 & \underline{92.5±4.5} \\
    \cmidrule(lr){1-5}
    \rowcolor{gray!20} \textbf{Ours} (th=0.5) & 92.7±1.6 & \textbf{98.9±1.3} & \underline{94.4±10.1} & \textbf{96.2±4.9} \\
    \rowcolor{gray!20} \textbf{Ours} (th=0.7) & \textbf{95.1±1.2} & \underline{97.3±2.1} & 85.4±11.7 & 92.1±8.5 \\
    \bottomrule
    \end{tabular}
    \end{threeparttable}
    \label{tab:performance_comparison_on_BAS}
\end{table}

\begin{table}[t]
    \setlength{\abovecaptionskip}{0pt}
    \setlength{\belowcaptionskip}{0pt}
    \caption{Performance comparison on EXACT’09 dataset. \\ Mean ± standard deviation is shown for each metric.}
    \centering
    \begin{threeparttable}
    \footnotesize
    \begin{tabular}{p{1.9cm}|ccC{1.0cm}C{1.0cm}}
    \toprule
    & Branch & Length (cm) & BD (\%) & TD (\%) \\
    \cmidrule(lr){1-5}
    Xu et al.~\cite{xu2015hybrid} & 128.7±60.3 & 94.8±44.7 & 51.7±10.8 & 44.5±9.4 \\
    Yun et al.~\cite{yun2019improvement} & 163.4±79.4 & 129.3±66.0 & 65.7±13.1 & 60.1±11.9 \\
    Qin et al.~\cite{qin2021learning} & 190.4 & 166.5 & 76.7±11.5 & 72.7±11.6 \\
    Zheng et al.~\cite{zheng2021alleviating} & 199.9 & 180.9 & 80.5±12.5 & 79.0±11.1 \\
    DTPDT~\cite{zhang2022differentiable} & \underline{203.9} & \underline{182.4} & \underline{82.1±10.6} & \underline{79.6±9.5} \\
    \cmidrule(lr){1-5}
    Neko \cite{balacey2013mise} & 84.5±40.5 & 61.9±30.9 & 35.5±8.2 & 30.4±7.4 \\
    UCCTeam \cite{nardelli2015optimizing} & 99.0±50.3 & 75.1±39.4 & 41.6±9.0 & 36.5±7.6 \\
    FF\_ITC \cite{inoue2013robust} & 198.3±98.6 & 177.1±97.0 & 79.6±13.5 & 79.9±12.1 \\
    MISLAB \cite{lo2012extraction} & 104.7±55.2 & 78.7±41.7 & 42.9±9.6 & 37.5±7.1 \\
    NTNU \cite{smistad2014gpu} & 72.4±37.8 & 54.3±33.9 & 31.3±10.4 & 27.4±9.6 \\
    \cmidrule(lr){1-5}
    \rowcolor{gray!20} \textbf{Ours} (th=0.5) & \textbf{219.3±65.5} & \textbf{196.6±53.9} & \textbf{88.3±24.3} & \textbf{85.6±20.0} \\
    \bottomrule
    \end{tabular}
    \end{threeparttable}
    \label{tab:performance_comparison_on_EXACT09}
\end{table}

\subsection{Implementation}
\subsubsection{Data preparation}
First, we stacked up the CT slices (in the DICOM format) to form 3D image data if needed. Then, thresholding was done to keep the Hounsfield Unit (HU) values within [-1000, 600]. The pixel values were further standardized to be within [0, 1]. During training, images were cropped into 32x128x128 cuboids owing to available GPU memory.

\subsubsection{Training Procedure}
We used PyTorch \cite{paszke2019pytorch} to implement our model. Data augmentation, including random flip, random affine, random blur, random noise, random motion, and random spike \cite{perez2021torchio}, was performed to expand the sample size. The model was trained on two NVIDIA GeForce RTX 2080 Ti cards with an Adam optimizer and a learning rate of $10^{-5}$. To implement our iterative training, we sampled 1000 cuboids with one of the strategies (e.g., \textit{l iter}) in Equation~\eqref{equ:training_strategy} and switched to the other strategy (e.g., \textit{h iter}) after finishing one epoch. The total number of epochs was determined based on the size of the training dataset. For example, on the BAS dataset, we trained the model for 100 epochs.

\subsection{Metrics}
\label{sec:experiments_metrics}
Ideally, the accuracy of model-based airway segmentation is best evaluated through the actual bronchoscopic procedure by experts. However, it is labor-intensive, and any additional examination that may prolong the clinical procedure should be avoided. Therefore, we only have \emph{reference} labels, instead of \emph{ground truth} labels. Hence, in cases of airway segmentation, the goal of a model is not to provide airway segments that perfectly match the reference but to recognize bronchioles as many as possible from CT images.

We adopted the Tree length Detected rate (TD) and Branch Detected rate (BD) from EXACT’09~\cite{lo2012extraction} to evaluate topological accuracy. Additionally, we employed Dice Similarity Coefficient (DSC, $\frac{2TP}{2TP+FP+FN}$) to evaluate the overall similarity between model predictions and the reference labels. Sensitivity ($\frac{TP}{TP+FN}$) was used to check the percentage of volumes the model detects in the reference labels.

\begin{table*}[t]
    \setlength{\abovecaptionskip}{0pt}
    \setlength{\belowcaptionskip}{0pt}
    \caption{Ablation study on BAS dataset. Mean ± standard deviation (\%) is shown for each metric.}
    \centering
    \begin{threeparttable}
    \begin{tabular}{cccccc|cccc}
    \toprule
    Backbone & $\mathfrak{L}_{pen}$ & $\mathfrak{L}_{ske}$ & Iter Train & Distill & Data Aug & DSC & Sensitivity & BD & TD \\
    \cmidrule(lr){1-10}
    $\checkmark$ & & & & & $\checkmark$ & 92.6±1.4 & 94.7±3.6 & 65.9±17.8 & 66.5±19.8 \\
    \rowcolor{gray!20} $\checkmark$ & $\checkmark$ & & & & $\checkmark$ & 92.0±1.6 & 95.2±3.9 & 71.4±14.7 & 77.7±17.8 \\
    $\checkmark$ & & $\checkmark$ & & & $\checkmark$ & 92.4±1.5 & 95.8±3.4 & 72.9±13.8 & 79.3±14.1 \\
    \rowcolor{gray!20} $\checkmark$ & $\checkmark$ & $\checkmark$ & & & $\checkmark$ & 91.7±1.4 & 96.0±2.9 & 75.2±13.5 & 81.9±12.8 \\
    $\checkmark$ & & & $\checkmark$ & & $\checkmark$ & 92.5±1.4 & 95.6±3.5 & 71.7±13.9 & 76.5±15.1 \\    
    \rowcolor{gray!20} $\checkmark$ & $\checkmark$ & $\checkmark$ & $\checkmark$ & & $\checkmark$ & 92.4±1.8 & 97.5±1.8 & 86.8±13.5 & 90.8±8.8 \\
    $\checkmark$ & $\checkmark$ & $\checkmark$ & $\checkmark$ & $\checkmark$ & & \textbf{93.1±1.5} & 97.7±1.8 & 90.8±9.6 & 92.9±4.6 \\
    \rowcolor{gray!20} $\checkmark$ & $\checkmark$ & $\checkmark$ & $\checkmark$ & $\checkmark$ & $\checkmark$ & 92.7±1.6 & \textbf{98.9±1.3} & \textbf{94.4±10.1} & \textbf{96.2±4.9} \\
    \bottomrule
    \end{tabular}
    \begin{tablenotes}
    \footnotesize
    \item ``Backbone'' represents the backbone model introduced in Section~\ref{sec:method_model}; $\mathfrak{L}_{pen}$ and $\mathfrak{L}_{ske}$ are Penalty Dice Loss and Skeleton Dice Loss proposed in Section~\ref{sec:method_loss_func}; ``Iter Train'' represents the iterative training strategy in Section~\ref{sec:method_training_strategy}; ``Distill'' represents the knowledge distillation module; ``Data Aug'' means the data augmentation applied to the training batches.
    \end{tablenotes}
    \end{threeparttable}
    \label{tab:ablation_study}
\end{table*}

\subsection{Performance comparison}
\begin{figure}[t]
  \centering
  \includegraphics[width=8cm]{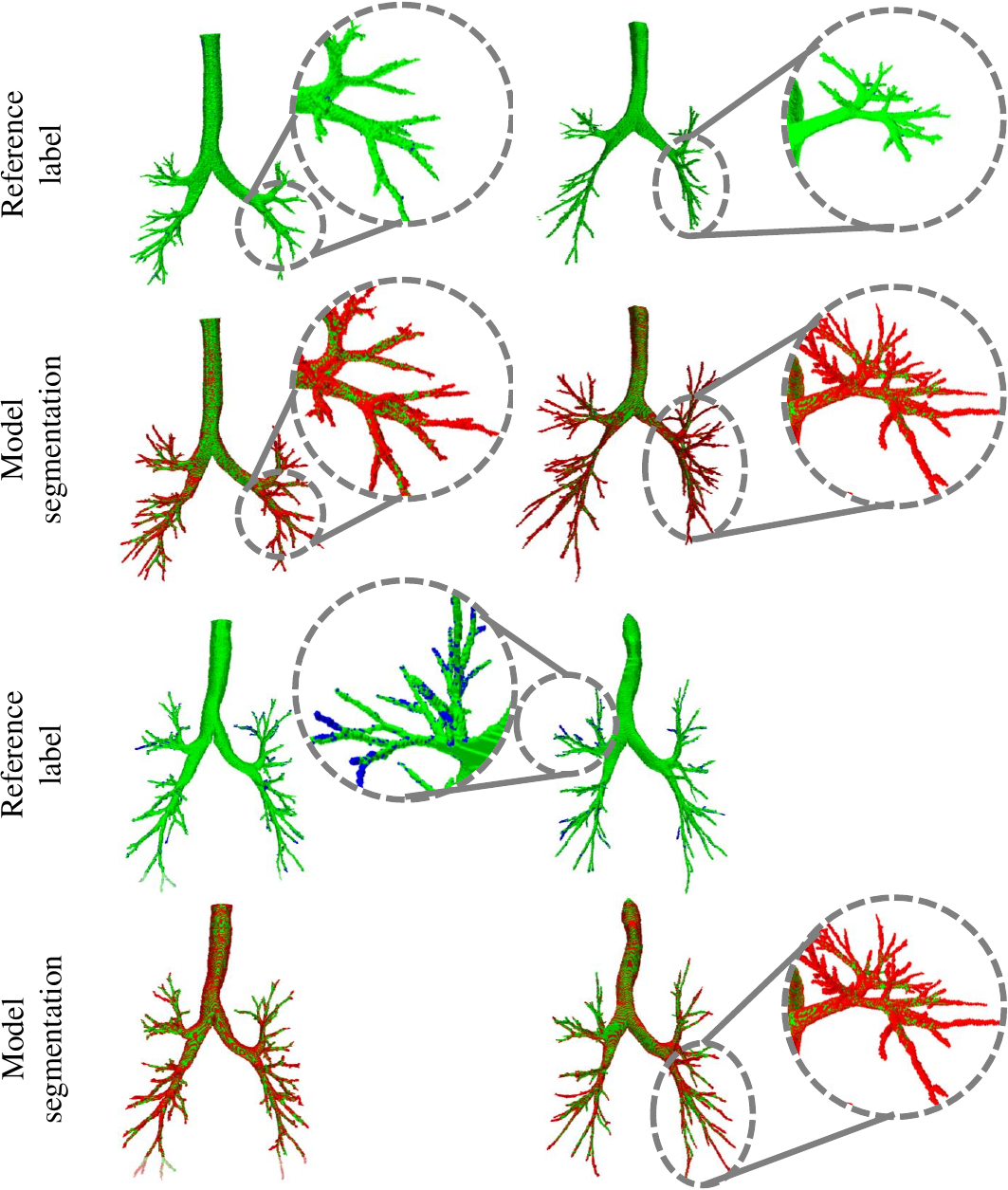}
  \caption{Exemplary test results on BAS and EXACT’09 datasets showing that NaviAirway achieves high topological accuracy while finding more finer bronchioles than reference labels. (\textbf{\textcolor{green}{Green}}: overlapped volume. \textbf{\textcolor{red}{Red}}: extra airway voxels found by the model. \textbf{\textcolor{blue}{Blue}}: missed airway voxels by the model.)
}
\label{fig:results}
\end{figure}

First, shown in Table~\ref{tab:performance_comparison_on_BAS} and~\ref{tab:performance_comparison_on_EXACT09},  we compared NaviAirway with existing methods~\cite{jin20173d, juarez2018automatic, schlemper2019attention, wang2019tubular, juarez2019joint, qin2019airwaynet, xue2020shape, qin2020learning, qin2021learning, zheng2021alleviating, xu2015hybrid, yun2019improvement, zhang2022differentiable, balacey2013mise, nardelli2015optimizing, inoue2013robust, lo2012extraction, smistad2014gpu} on both BAS~\cite{qin2020learning} and EXACT’09~\cite{lo2012extraction} datasets. For the BAS dataset (Table~\ref{tab:performance_comparison_on_BAS}), our method outperformed others in both topological accuracy metrics (BD: 94.4±10.1 and TD: 96.2±4.9) and Sensitivity (98.9±1.3). Note that ``th'' means the threshold value to decide whether a voxel in the prediction map belongs to the airway or not. Here we show the results of two threshold values, 0.5 and 0.7. As shown in Figure~\ref{fig:results}, when th=0.5, our NaviAirway detects more bronchioles which are not shown in the reference label and are previously regarded as ``false positive'' predictions. However, as mentioned in Section~\ref{sec:experiments_metrics}, the reference labels may miss some airways due to the limitation of manual labeling. After the exam by a bronchologist, these ``extra bronchioles'' are considered to be true. That is why ``th=0.5'' has a lower DSC value than ``th=0.7''. A lower DSC does not necessarily mean the performance is worse as there might be missed airways in the reference labels. On average, our method detects bronchioles up to the 12th generation, whereas the mean and median values of the detected generation number are 7.9 and 7.5, respectively.

Table~\ref{tab:performance_comparison_on_EXACT09} shows the performance comparison on the EXACT’09 dataset. In this table, ``Branch'' represents the number of airway branches and ``Length'' is detected airway length in cm. Our NaviAirway detects the longest airway (196.6±53.9) and up to the 13th generation on average while having the highest BD (88.3±24.3) and TD (85.6±20.0).

Moreover, we also tested our method in the ATM22 challenge~\cite{zheng2021alleviating, zhang2021fda, yu2022break, qin2019airwaynet}. Among the submissions in the long-term validation phase, NaviAirway achieved the third highest BD (95.5) and the second highest TD (96.3). Note that we adopted the same set of hyperparameters for the training of the three datasets (BAS, EXACT’09, and ATM22).

\subsection{Ablation study}
We conducted all ablation studies on the BAS dataset to evaluate the contribution of each component in our proposed model to performance improvement. Additionally, we tested various settings of hyperparameters in each component to further optimize the model's performance.

\subsubsection{The effectiveness of bronchiole-sensitive loss function} Step by step, we verified the effectiveness of NaviAirway. As shown in Table~\ref{tab:ablation_study}, compared with the baseline case (``Backbone''), Penalty Dice Loss $\mathfrak{L}_{pen}$ alone boosted the performance from 65.9 BD and 66.5 TD to 71.4 (+5.5) BD and 77.7 (+11.2) TD, and Skeleton Dice Loss $\mathfrak{L}_{ske}$ alone boosted the performance from 65.9 BD and 66.5 TD to 72.9 (+7.0) BD and 79.3 (+12.8) TD. When combining $\mathfrak{L}_{pen}$ and $\mathfrak{L}_{ske}$, the accuracy increased to 75.2 (+9.3) BD and 81.9 (+15.4) TD.

\subsubsection{The effectiveness of iterative training} The iterative training strategy alone (see the row of ``Backbone'' + ``Iter Train'' in Table~\ref{tab:ablation_study}) improved the baseline model by +5.8 BD and +10.0 TD, respectively. When combining iteration training and proposed loss functions, the performance was further boosted to 86.8 (+20.9) BD and 90.8 (+24.2) TD.

\subsubsection{The effectiveness of knowledge distillation from unlabeled images} On top of $\mathfrak{L}_{pen}$, $\mathfrak{L}_{ske}$, and the proposed iterative training strategy, knowledge distillation from unlabeled images could further increase the model performance to 94.4 (+7.6) BD and 96.2 (+5.4) TD. Additionally, comparing the last two rows, we found that data augmentation only had a minor effect on performance improvement. This indicates that the performance gain mainly comes from our proposed modules.

\begin{table}[h]
    \setlength{\abovecaptionskip}{0pt}
    \setlength{\belowcaptionskip}{0pt}
    \caption{Study on hyperparameter selection of the loss function. \\ Mean ± standard deviation (\%) is shown for each metric.}
    \centering
    \begin{threeparttable}
    \footnotesize
    \begin{tabular}{l|cccc}
    \toprule
    & DSC & Sensitivity & BD & TD \\
    \cmidrule(lr){1-5}
    $\alpha = 3$ & 93.2±2.1 & 97.3±1.5 & 92.5±9.1 & 94.3±5.3 \\
    $ker = 5$ & 92.9±1.4 & 98.2±1.6 & 94.6±10.9 & 95.9±5.0 \\
    $J = 1$ & 93.6±2.0 & 98.0±1.2 & 93.9±9.8 & 94.9±5.1 \\
    $J = 2$ & 93.0±1.5 & 98.4±1.4 & 94.3±10.0 & 96.4±4.9 \\
    $J = 4$ & 92.6±1.6 & 98.8±1.2 & 94.2±10.2 & 96.2±5.0 \\
    \rowcolor{gray!20} \textbf{Default}\tnote{1} & 92.7±1.6 & 98.9±1.3 & 94.4±10.1 & 96.2±4.9 \\
    \bottomrule
    \end{tabular}
    \begin{tablenotes}
    \footnotesize
    \item[1] ``Default'' represents our default setting where $\alpha = 2$, $ker = 3$, and $J = 3$. Other rows represent that only one hyperparameter varies at a time (e.g., $\alpha = 3$ represents the hyperparameter setting is $\alpha = 3$, $ker = 3$, and $J = 3$).
    \end{tablenotes}
    \end{threeparttable}
    \label{tab:loss_func_hyperparameters}
\end{table}

\subsubsection{Modulating term, kernel size, and number of iterations in loss function} For Penalty Dice Loss $\mathfrak{L}_{pen}$, we used $\alpha$ to control the modulating effect. Table~\ref{tab:loss_func_hyperparameters} shows the two cases of $\alpha$: 2 and 3. We set $\alpha = 2$ by default. Besides, for Skeleton Dice Loss, the kernel size $ker$ and the number of iteration(s) $J$ (see Algorithm~\ref{algo:algorithm_skeleton}) determine the quality of skeletonization. We tested the cases where $ker$ is 3 or 5 and $J$ ranged from 1 to 4. We finally set $ker = 3$ and $J = 3$ because this setting led to satisfactory performance and computational efficiency.

\begin{table}[h]
    \setlength{\abovecaptionskip}{0pt}
    \setlength{\belowcaptionskip}{0pt}
    \caption{Comparison of different training strategies.}
    \centering
    \begin{threeparttable}
    \scriptsize
    \begin{tabular}{lc|lc}
    \toprule
    Strategy & DSC (\%) & Strategy & DSC (\%) \\
    \cmidrule(lr){1-4}
    $\mbox{\textit{l\ iter}}$ only ($\tau_{l}=1$) & 89.6 & First $\mbox{\textit{l\ iter}}$ then $\mbox{\textit{h\ iter}}$ & 90.0 \\
    $\mbox{\textit{l\ iter}}$ only ($\tau_{l}=0.1$) & 90.4 & First $\mbox{\textit{h\ iter}}$ then $\mbox{\textit{l\ iter}}$ & 88.2 \\
    $\mbox{\textit{h\ iter}}$ only ($\tau_{h}=1$) & 87.3 & Same frequency & 92.6 \\
    $\mbox{\textit{h\ iter}}$ only ($\tau_{h}=0.1$) & $\sim$0 & Iterative & 95.1 \\
    \bottomrule
    \end{tabular}
    \end{threeparttable}
    \label{tab:different_training_strategies}
\end{table}

\subsubsection{Comparison of training strategies} In Table~\ref{tab:different_training_strategies}, we investigated different $\beta$ values and compared the four training strategies (the right two columns) with the threshold being 0.7. Results show that our iterative training strategy performs the best.


\begin{table}[t]
    \setlength{\abovecaptionskip}{0pt}
    \setlength{\belowcaptionskip}{0pt}
    \caption{Generality of NaviAirway (tested on BAS dataset).
    \\ Mean ± standard deviation (\%) is shown for each metric.}
    \footnotesize
    \centering
    \begin{tabular}{l|cccc}
    \toprule
    & DSC & Sensitivity & BD & TD \\
    \cmidrule(lr){1-5}
    3D U-Net~\cite{cciccek20163d} & 92.9±1.7 & 95.8±2.3 & 66.5±18.8 & 72.3±18.8 \\
    w/ NaviAirway & 90.5±1.5 & 97.3±2.2 & 81.0±12.2 & 85.5±10.5 \\
    \cmidrule(lr){1-5}
    V-Net~\cite{milletari2016v}& 85.9±3.4 & 81.8±7.0 & 34.2±9.1 & 35.0±9.8 \\
    w/ NaviAirway & 87.3±2.5 & 83.6±6.5 & 67.8±12.5 & 74.6±8.7 \\
    \cmidrule(lr){1-5}
    VoxResNet~\cite{chen2018voxresnet} & 85.8±6.3 & 78.3±9.8 & 29.8±9.9 & 33.1±10.2 \\
    w/ NaviAirway & 88.7±4.6 & 83.0±7.2 & 70.3±14.8 & 77.7±10.1 \\
    \cmidrule(lr){1-5}
    Wang et al.~\cite{wang2019tubular} & 93.5±2.2 & 88.6±8.8 & 93.4±8.0 & 85.6±9.9 \\
    w/ NaviAirway & 92.1±3.7 & 90.0±8.3 & 93.8±7.5 & 91.9±8.8 \\
    \bottomrule
    \end{tabular}
    \label{tab:generality}
\end{table}

\begin{table}[t]
    \setlength{\abovecaptionskip}{0pt}
    \setlength{\belowcaptionskip}{0pt}
    \caption{Performance on unseen private dataset (QMH) \\ Mean ± standard deviation (\%) is shown for each metric.}
    \centering
    \begin{threeparttable}
    \footnotesize
    \begin{tabular}{cccc}
    \toprule
    DSC & Sensitivity & BD & TD \\
    \cmidrule(lr){1-4}
    86.3±4.9 & 95.5±1.3 & 90.6±8.2 & 84.3±25.3 \\
    \bottomrule
    \end{tabular}
    \end{threeparttable}
    \label{tab:QMH}
\end{table}

\subsection{Method generality}
Our NaviAirway method can be generalized to other backbone models. As shown in Table~\ref{tab:generality}, we simply applied NaviAirway to 3D U-Net~\cite{cciccek20163d}, V-Net~\cite{milletari2016v}, VoxResNet~\cite{chen2018voxresnet}, and Wang et al.~\cite{wang2019tubular} without careful hyperparameter tuning. The Results show that NaviAirway boosts the four models by a considerable margin. Additionally, Table~\ref{tab:QMH} examines the performance of our model in the unseen dataset (QMH). NaviAirway still achieves high accuracy (Sensitivity: 95.4, BD: 90.2, and TD: 83.7), indicating our method is robust to new data.

\section{Discussion}

\subsubsection{Tackling the size imbalance problem by sampling techniques}
\label{sec:discussion_training}
In Section~\ref{sec:method_training_strategy}, we devised our strategy based on the ratio $r_i^j$, which can be interpreted as the ratio of surface area to volume of a given airway branch of interest. Since airway branches are long tubular structures, $r_i^j$ can be approximated as $r_i^j \approx \frac{2 \pi \hat{r} \hat{l}}{\pi \hat{r}^{2} \hat{l}} = \frac{2}{\hat{r}}$, where $\hat{r}$ denotes the average radius and $\hat{l}$ denotes the average branch length. The size imbalance problem arises from the uneven distribution of airway branch volumes. The size of low-generation airways is approximately $(\frac{\hat{r}_{l}}{\hat{r}_{h}})^{2}$ (where subscript $l$ means low and subscript $h$ means high) times larger than those of high generations. Therefore, we use $r_i^j$ to down- or up-sample the CT image cuboids. There are various approaches to sampling theoretically, and we presented an effective approach. In future work, we plan to conduct more in-depth investigations and explore alternative strategies.

\subsubsection{Interpreting Penalty Dice Loss}
\label{sec:discussion_penalty_dice_loss}
\begin{figure}[h]
  \centering
  \includegraphics[width=5.5cm]{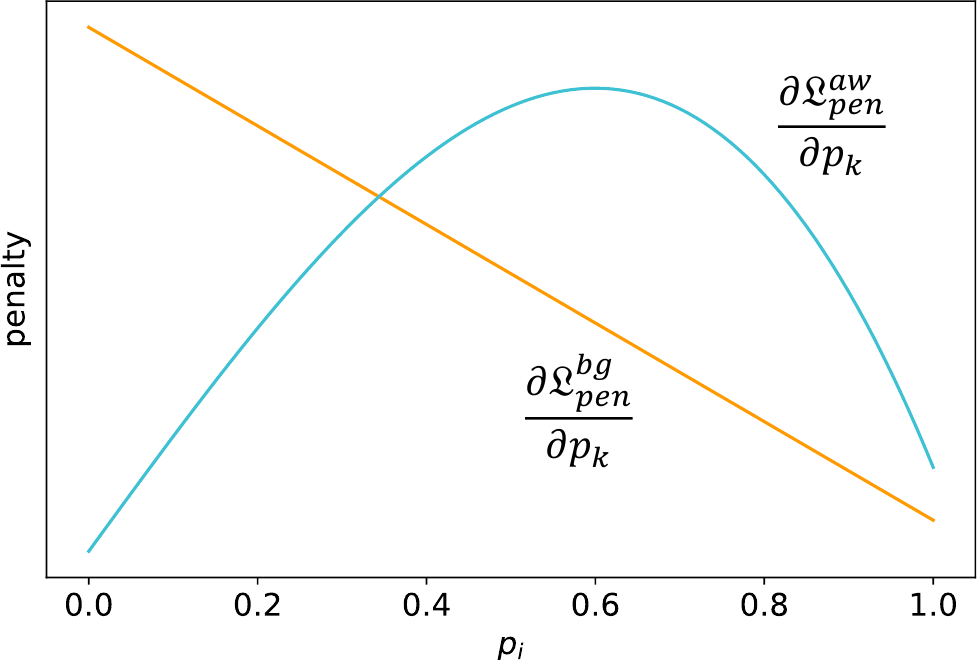}
  \caption{The changes of $\frac{\partial \mathfrak{L}_{pen}^{aw}}{\partial p_k}$ and $\frac{\partial \mathfrak{L}_{pen}^{bg}}{\partial p_k}$ with different $p_k$ values. The blue curve shows that $\mathfrak{L}_{pen}^{aw}$ penalizes heavily on airway segments of moderate confidence and rejects those of too low confidence. }
\label{fig:penalty_dice_loss}
\end{figure}

To better understand the proposed loss function, we can analyze the derivative of $\mathfrak{L}_{pen}$ (Equation~\eqref{equ:penalty_dice_loss_func}). For $p_k$ on the airway prediction map, $\frac{\partial \mathfrak{L}_{pen}}{\partial p_k}=\frac{\partial \mathfrak{L}_{pen}^{aw}}{\partial p_k}$, while on the background prediction map, $\frac{\partial \mathfrak{L}_{pen}}{\partial p_k}=\frac{\partial \mathfrak{L}_{pen}^{bg}}{\partial p_k}$. As shown in Figure~\ref{fig:penalty_dice_loss}, $\frac{\partial \mathfrak{L}_{pen}^{bg}}{\partial p_k}$ follows a linear pattern, whereas $\frac{\partial \mathfrak{L}_{pen}^{aw}}{\partial p_k}$ indicates that $\mathfrak{L}_{pen}^{aw}$ penalizes heavily on airway segments of moderate confidence (which could be bronchioles) and rejects those of too low confidence (which tend to be noises).



\subsubsection{Complementary Metrics Needed}
\label{sec:discussion_complementary_metrics}
Most existing works followed the metrics built in the EXACT‘09 challenge in 2009 \cite{lo2012extraction}. However, because we only have the ``reference'' labels, instead of the ``ground truth'' segmentation, complementary metrics are needed. First, quantitative evaluation was insufficient. We proposed to include visual inspection because some ``False Positive ($FP$) airway segments'' might be true airway branches that were missed by the ``reference'' labels. As shown in Table~\ref{tab:discussion_review_metrics}, we refine Branch Detected (BD) and Tree-length Detected (TD) to be the ratio between ``True Positive ($TP$) airway segments'' plus true airway segments missed by ``reference'' labels over the segments in ``reference'' labels and denote the two adjusted metrics as Adjusted Branch Detected (ABD) and Adjusted Tree-length Detected (ATD). Results show that NaviAirway helped the model learn the general features of airways, so the model could detect more finer bronchioles that were not shown in the ``reference'' labels.

\begin{table}[t]
    \setlength{\abovecaptionskip}{0pt}
    \setlength{\belowcaptionskip}{0pt}
    \caption{Review metrics for airway segmentation. \\ Mean ± standard deviation (\%) is shown for each metric.}
    \centering
    \begin{threeparttable}
    \footnotesize
    \begin{tabular}{lcc}
    \toprule
    Dataset & ABD\tnote{1} & ATD\tnote{2} \\
    \cmidrule(lr){1-3}
    BAS & 114.2±18.1 & 113.2±17.5 \\
    QMH & 132.4±24.5 & 108.6±13.5 \\
    \bottomrule
    \end{tabular}
    \begin{tablenotes}
    \footnotesize
    \item[1] ABD: Adjusted Branch Detected.
    \item[2] ATD: Adjusted Tree-length Detected.
    \end{tablenotes}
    \end{threeparttable}
    \label{tab:discussion_review_metrics}
\end{table}

\section{Conclusions}

In this paper,  we present a novel airway segmentation pipeline that provides extensive airway road maps with more detailed bronchioles. This is achieved through our proposed bronchiole-sensitive loss function for airway topology preservation and an iterative training strategy to address the size imbalance problem. Additionally, we leverage unlabeled chest CT images to distill airway branch features using a teacher-student training framework. Our approach is robust and outperforms existing methods on new CT scans from different systems and institutions. Furthermore, our method is compatible with various backbone models, thus improving their performance. Beyond airway segmentation, our approach can be extended to segmenting other fine and long tubular structures in biomedical images.
\section*{Acknowledgments}


{\appendices
}


\bibliographystyle{IEEEtran}
\bibliography{IEEEabrv,references}

 
%




\newpage
 





\end{document}